# Publish or impoverish: An investigation of the monetary reward system of science in China (1999-2016)

Wei Quan, School of Information Management, Wuhan University, Wuhan, China

Bikun Chen, School of Economics and Management, Nanjing University of Science and Technology, Nanjing, China

Fei Shu, School of Information Studies, McGill University, Montreal, Canada

Abstract
**Purpose** – The purpose of this study is to present the landscape of the cash-per-publication reward policy in China and reveal its trend since the late 1990s.
**Design/methodology/approach** – This study is based on the analysis of 168 university documents regarding the cash-per-publication reward policy at 100 Chinese universities.
**Findings** – Chinese universities offer cash rewards from 30 to 165,000 USD for papers published in journals indexed by Web of Science (WoS), and the average reward amount has been increasing for the past 10 years.
**Originality/value** – The cash-per-publication reward policy in China has never been systematically studied and investigated before except for in some case studies. This is the first paper that reveals the landscape of the cash-per-publication reward policy in China.
**Paper type** Research paper

Introduction
Although monetary rewards have been used for recognizing scientific achievement since the eighteen century, it is not regarded as the major reward system in science (Robert King Merton, 1973), in which scientists try to publish their works and receive the recognition of their peers as the reward. Since academic prizes consisting of cash rewards are awarded only to very few scientific elites, they are considered as the metaphors of *prestige* rather than simply large sums of money (Zuckerman, 1992). However, the reward system in science changed when the monetary reward incentive for publication was introduced in 1980s. It is reported that this incentive can promote research productivity (Franzoni, Scellato, & Stephan, 2011) but might create a negative goal displacement effect (Frey, Osterloh, & Homberg, 2013; Osterloh & Frey, 2014).

Since the early 1990s, Chinese research institutions have initiated the cash-per-publication reward polices in which Chinese scholars could get cash for each eligible publication. The purpose of publishing their works is not only to advance knowledge and win recognition, but also to earn cash (Sun & Zhang, 2010; L. Wang, 2016). Since these cash-per-publication reward policies vary by institution and some policies are internal or confidential, they have never been systematically investigated except for in some case studies. The purpose of this study is to present the landscape of the cash-per-publication reward policy in China[1] and reveal its trend since the late 1990s.

*China's Scientific Activity*

With the significant development of China's economy, China's scientific activity is experiencing a period of rapid growth. As Figure 1 shows, China's scientific research inputs and outputs have exhibited a consistent growth pattern over the past 20 years; from 1995 to 2013, China's expenditures on Research and Development (R&D) increased almost 33 times from 5.23 billion USD[2] to 177.70 billion USD, while its number of international publications (indexed by Web of Science) increased about 17 times from 13,134 to 232,070 (National Bureau of Statistics of China, 1996-2014). Indeed, China, the second largest share of international scientific production per country since 2009, contributes 16.3% of scientific articles indexed by Web of Science (hereafter referred to as WoS) (Institute of Scientific and Technical Information of China, 2016).

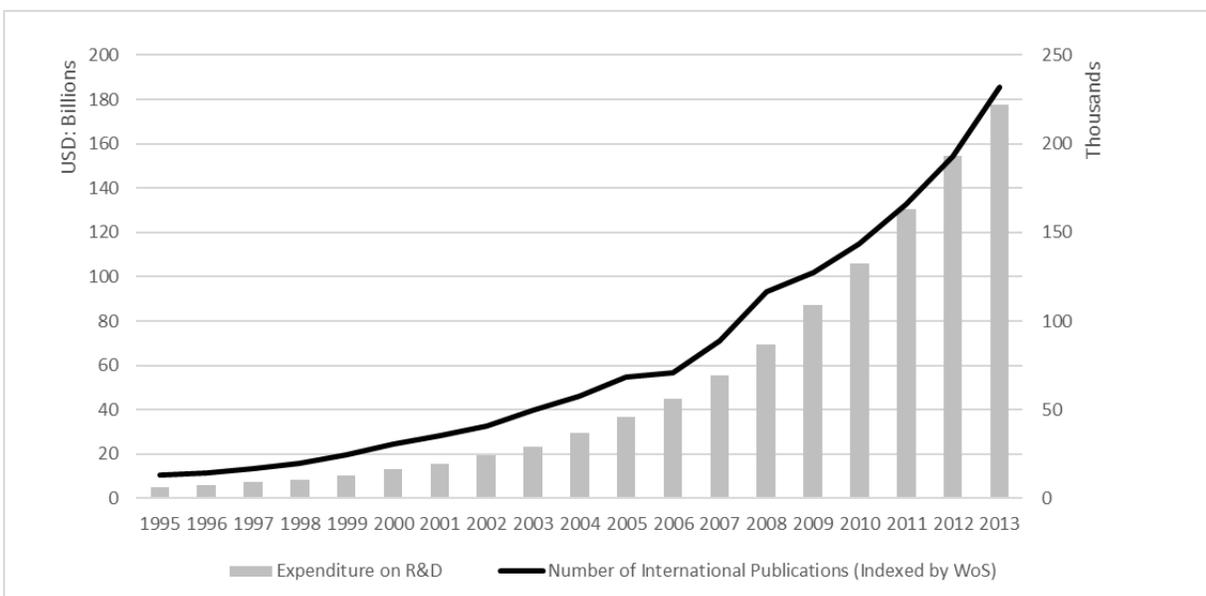

**Figure 1 Research Inputs and Outputs in China (1995-2013)**

*Universities in China*

In China, although the universities, the research institutions, the enterprises, the hospitals not affiliated with universities, and other sectors are all involved in scientific activity, the universities play the dominant role in China's scientific research output, contributing 82.8% of monographs and 73.4% of journal articles, including 83.0% of WoS papers (National Bureau of Statistics of China, 2015). There are 2,595 higher education institutions in China, including 1,236 universities offering undergraduate programs (Ministry of Education of China, 2016).

---

[1] In this study, China refers to the mainland China, which is the geopolitical area under the direct jurisdiction of the People's Republic of China excluding Hong Kong and Macau.

[2] In this study, all cash values in Chinese Yuan (CNY) were converted to US dollar (USD) at the rate 1 CNY = 0.15 USD. The exchange rate was retrieved from xe.com on Oct. 3, 2016.

Traditionally these universities vary by ownership, speciality, and region; however, they also can be classified into three tiers by two national research programs: *Project 211* and *Project 985*.

Project 211 was initiated in 1995 by China's Ministry of Education. The objective of this project was to construct 100 world-class universities in the beginning of the twenty-first century (Ministry of Education of China, 2000). The Chinese government offers preferential policies and financial support to designated universities who are admitted to this project, and has contributed around 2.7 billion USD to it (Tang & Yang, 2008). Eventually, 116 universities were admitted to Project 211, forming an elite group of universities occupying 70% of national research funding and supervising 80% of doctoral students (Tang & Yang, 2008). Today, 112 universities are still included in Project 211, even after 4 universities merged.

Project 985 was first announced by Zemin Jiang, former Chairman of the People's Republic of China, on May 4, 1998 to promote the development of a Chinese equivalent of the US Ivy League (Chen, 2006). This *Chinese Ivy League* started with 9 universities in 2009 and accepted another 30 universities in the following two years. These 39 universities are all Project 211 universities, but receive more government funding than other Project 211 universities (Mohrman, 2005). Both Project 985 and Project 211 ceased admission in 2011, grouping Chinese universities into a 3-tier pyramid hierarchy as shown in Figure 2: 39 universities within Project 985 (hereafter referred to as *Tier 1* universities), 73 universities within Project 211 but excluded from Project 985 (hereafter referred to as *Tier 2* universities), and 1,124 other universities (hereafter referred to as *Tier 3* universities).

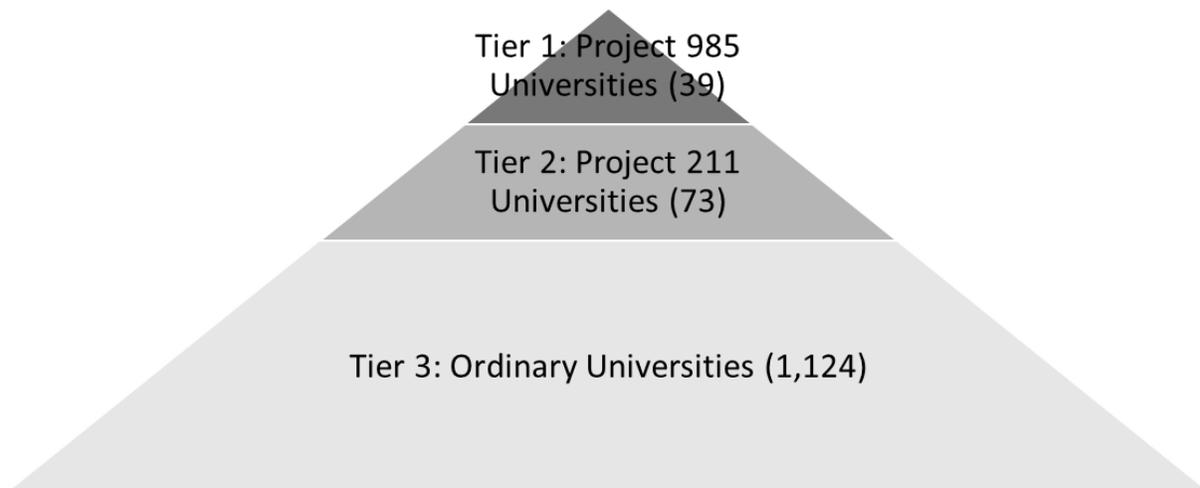

**Figure 2 The Pyramid Hierarchy of Chinese Universities**

To construct world-class universities, the Chinese government differentiates universities and allocates most funding to a few elite universities, which lead to the "Matthew Effect" among Chinese universities. Figure 3 presents a huge gap between the elite (Tier 1 and Tier 2) and ordinary (Tier 3) universities in terms of their average annual budgets. From 2002 to 2015, the average annual budget of Tier 1 and Tier 2 universities increased from 23.86 million USD to

113.05 million USD while the mean budget of Tier 3 universities increased from 1.89 million USD to 9.27 million USD. Tier 1 and Tier 2 universities' budget is, on average, 12 times more than Tier 3 universities budget at all times (Ministry of Education of China, 2003-2016). Although a new national research program, "Double World-Class", was announced by the Chinese government in 2016, it was not in effect during the period of our investigation.

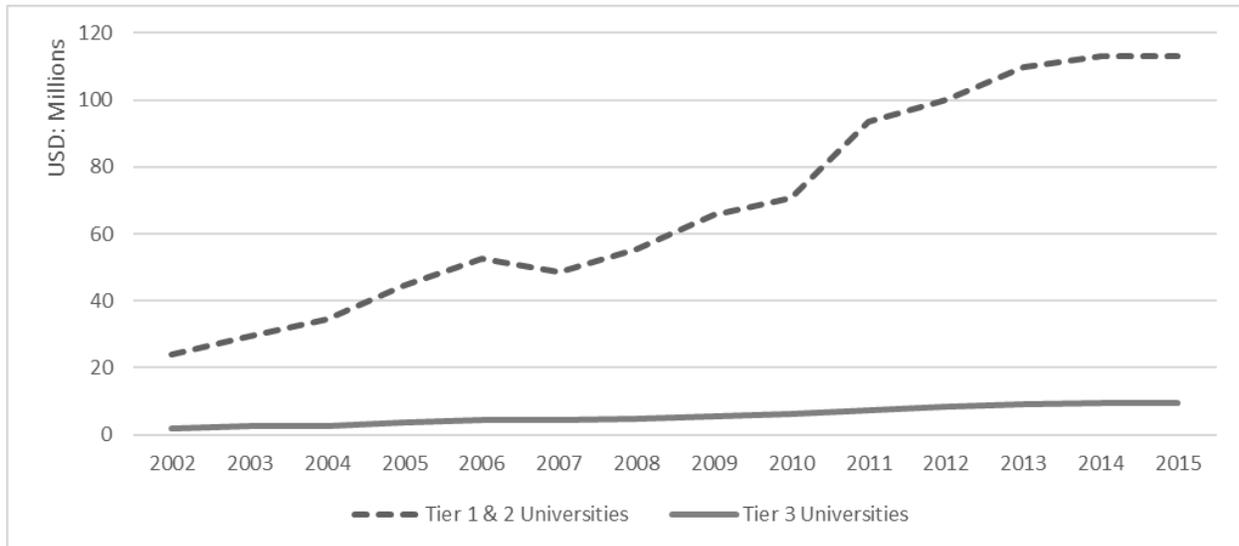

**Figure 3 Comparison of Average University Budget between Elite Universities and Ordinary Universities (2002-2014)**

*The Cash-per-publication Reward Policy in China*

Since the 1980s, to increase the international visibility of Chinese research, the number of WoS papers has been used to evaluate the research performance in China of both institutions and individuals (Gong & Qu, 2010). Chinese scholars are required to publish WoS papers to attain promotion, while their affiliated institutions need the number of WoS papers for ranking and funding application (Y. Wang & Li, 2015). Chinese universities and research institutions also offer preferential policies and monetary rewards to encourage their scholars to publish in journals indexed by WoS (Peng, 2011).

The first cash-per-publication reward policy (hereafter referred to as the cash reward policy) was launched by the Department of Physics at Nanjing University around 1990. Initially, scholars received 25 USD for each WoS paper, and the amount increased to between 60 and 120 USD in the mid-1990s (Swinbanks, Nathan, & Triendl, 1997). As the first to apply the WoS to research evaluation, Nanjing University topped the list of Chinese universities publishing most WoS papers seven years in a row in the 1990s (Gong & Qu, 2010); its research evaluation policy and cash reward policy were then copied by other universities and research institutions. Today, every university and research institution in China has established their own cash reward policies.

*Regional Difference in China*

China consists of 31 provincial-level divisions that were traditionally grouped into seven geographical regions: North, Northeast, Northwest, Center, East, Southwest, and South. Economic development in different regions differs significantly. The GDP per capita in the developed regions (i.e. the North, Northeast, East and South) are much higher than those in the developing regions (i.e. the Northwest, Center, and Southwest) as shown in Table 1.

Since Chinese universities are financially supported by not only the central government but also the local government, regional difference in economic development may lead to differences in the financial capacities of Chinese universities from different regions. Universities in developed regions may have adequate budgets, offering greater monetary reward compared to universities in developing regions. As Table 1 indicates, the average university budgets in the developed regions are much higher than those in the developing regions.

**Table 1 GDP per capita and Average University Budget in China by Region in USD (2014)**

|  | North | Northeast | Northwest | Center | East | Southwest | South |
|---|---|---|---|---|---|---|---|
| **GDP per capita** | $8,457.28 | $7,853.83 | $5,881.87 | $6,095.42 | $9,544.21 | $5,017.62 | $7,965.03 |
| **Avg. University Budget (in millions)** | $29.63 | $17.67 | $11.58 | $11.16 | $18.05 | $9.69 | $15.47 |

Source: National Bureau of Statistics of China (2015). *China Statistical Yearbook*. Beijing, China Statistics Press. Ministry of Education of China (2015). *Scientific Statistics in Higher Education Institutions - 2015*. Beijing, Higher Education Press.

Literature Review

*Reward System of Science*

Robert King Merton (1973) presents the *sociology* of science with a rewards system and recognition model. He states that science could be regarded as a social institution, with values, norms, and organization (Robert King Merton, 1957, 1973). This institution can reward its members (scientists) for their performance. Members also would like to present their achievements in order to get the rewards (Robert King Merton, 1957). Robert King Merton (1973) also points out that rewards for scientific achievement can be given only if others recognize it. As a result, scientists are eager to publish their works; peers read the publications and recognize their achievements by citing or acknowledging them in their own works. Based on Merton's recognition model (Robert King Merton, 1973), the reward system of science is also described as a *reward triangle* consisting of authorship, citations, and acknowledgements (Cronin & Weaver-Wozniak, 1993).

Previous studies suggest that other forms of recognition in addition to the "reward triangle" should be added to the reward system of science. Blume and Sinclair (1973) point out that academic prizes, honorary fellowships, and service on academic committees should be recognized as academic achievement. Sugimoto, Russell, Meho, and Marchionini (2008) compare citation counts and academic mentoring impact, and indicate that academic mentorship should also be granted recognition for its contribution to the spread of knowledge. As social media and other forms of dissemination are being incorporated into scientific practices, a multifaceted reward system has been identified that includes social media mentions, readership counts, and so on (Desrochers et al., 2015).

*Academic Monetary Rewards in History*

In 1719, the first academic prize was initiated by *Académie des Sciences* in France to award scientists who contribute to the advancement of knowledge in Astronomy. Thereafter, some academic prizes, with or without cash incentives, were introduced by *Académie des Sciences* and *Royal Society of London* to reward new scientific findings or past accomplishments. The establishment of The Nobel Prize, the largest monetary prize in the academic world, turned the academic prize into a metaphor for accomplishment and prestige (Zuckerman, 1977).There are now various academic prizes with big monetary rewards recognizing academic achievement locally and internationally. These large awards reshape the reward system of science's upper reaches (Zuckerman, 1992).

However, since these rich academic prizes are awarded only to very few outstanding scholars, they are valued on the basis of their representations of honors rather than their cash values (Zuckerman, 1992). In addition, Robert King Merton (1957) even claims that winning the monetary reward should not be the main purpose of any scientific activity because it may break the norm of *disinterestedness,* referring to rewards for action unaffected by self-interest. He states that:

> Like other institutions also, science has its system of allocating rewards for performance of roles. These rewards are largely honorific, since even today, when science is largely professionalized; the pursuit of science is culturally defined as being primarily a disinterested search for truth and only secondarily a means of earning a livelihood. (Robert King Merton, 1957, p. 659)

In 1986, the UK adopted the Research Assessment Exercise (RAE), which allocates national funds to departments based on past performance and peer review. The monetary incentives to publish, regarded as a reform, spread over the world afterwards; some countries even introduced a system of cash bonuses to individuals rather than institutions for each article published in top international scientific journals (Franzoni et al., 2011). Indeed, the economic incentives affect the university research at both the institution (Thursby, Jensen, & Thursby, 2001; Thursby & Thursby, 2002) and the individual levels (Frey et al., 2013; Osterloh & Frey, 2014).

*The Monetary Reward in China*

As described above, the cash reward policy was launched in China to promote scientific productivity and publication. Through case studies, some Chinese scholars found that monetary rewards could increase scholars' motivation and improve productivity in publishing WoS papers (Z Li & Zhang, 2008; ZW Li & Zhong, 2013; Shan, Han, & Zhao, 2013; Zeng, An, & Wang, 2012).

However, no study compares the cash reward policies in different universities nor presents the landscape of the monetary reward for publications in China.

Cash reward policies also create some negative effects. Chinese researchers may favour fast research that leads to quick, cashable publications as opposed to long-term research; in essence, publishing in WoS journals can become the only research goal (Jin & Rousseau, 2004). Some Chinese scientists may resort to plagiarized or fabricated research, purchase ghostwritten papers, or sell authorship (Hvistendahl, 2013; Qiu, 2010). Monetary reward also amplifies the existing "Matthew Effect" among Chinese universities (Zhong & Chen, 2008). Compared with Tier 3 universities, Tier 1 and Tier 2 universities dominate scientific resources, and thus could offer more cash rewards for publications, motivating their scholars to produce more (J. Li, 2013; Qi, 2009).

Since the monetary reward is an internal award, it is only announced by Chinese universities via internal documents. Some universities even keep it *confidential* to avoid competition from other universities. Although the cash reward policy has been applied for 20 years, we still know little about 1) the range of amounts paid to individuals for publications; 2) if the cash award varies with the quality of journals; 3) if the cash reward policy differs significantly from one university to another. The purpose of this study is to present the landscape of the cash-per-publication reward policy in China and address these questions.

Please note that the scope of this study is limited to research in natural science, including engineering and medical science. Since social science and humanities have more localized interests in research and varied methods of disseminating knowledge, beyond journal publications, most Chinese universities have different systems for evaluating and awarding research performance in social science and the humanities. For Chinese scholars in social science and humanities, there are additional approaches to winning cash rewards, such as publishing articles in domestic journals or publishing monographs. On the other hand, the cash rewards are only applied to publications in WoS journals in natural science.

## Methodology

*Data Collection*

In order to present the landscape of the cash-per-publication reward policy in China, we sampled 100 Chinese universities and investigated their cash reward policies since the 1990s. Both stratified sampling and convenience sampling were used.

First, considering the 3-tier pyramid hierarchy of Chinese universities and regional differences, we classified all 1,236 Chinese universities into 21 categories by tiers and regions. Second, we tried to retrieve the cash reward policies from universities in each category to ensure that the sample is representative. Since most cash reward policies are recorded in internal documents that may not be externally accessible, we had to select universities from each category based on data availability. We used the Chinese search engine *Baidu* to locate such information and retrieved it from the official websites of each selected university[3]. Finally, a manual validation was conducted to ensure that the retrieved documents were official and valid.

As Table 2 shows, 100 Chinese universities were selected for the investigation: 25 universities in Tier 1, 33 universities in Tier 2, and 42 universities in Tier 3. The samples also represent Chinese universities from all seven regions in China, as discussed above. Since some Chinese universities had multiple cash reward policies (e.g., modified or new ones), two or more cash reward policies were found in some universities during the period of the investigation. Eventually, 168 cash reward policies were retrieved from these 100 universities. 45 universities contributed one policy each, while 45 universities contributed two; Zhejiang University and Guizhou Normal University issued five and four cash reward policies, respectively, while 8 universities contributed three each. The first cash reward policy that we found was issued in 1999; the number of cash reward policies increased afterwards and reached its peak of 21 in 2015. Eight policies were even issued in 2016 as we started this investigation.

**Table 2 Distribution of the Sample Universities by Tier and Region**

|        | North    | Northeast | Northwest | Center   | East     | Southwest | South   | Total      |
|--------|----------|-----------|-----------|----------|----------|-----------|---------|------------|
| **Tier 1** | 3(10)    | 3(4)      | 3(4)      | 4(5)     | 7(11)    | 3(3)      | 2(2)    | 25(39)     |
| **Tier 2** | 4(19)    | 3(7)      | 4(9)      | 4(7)     | 12(20)   | 3(7)      | 3(4)    | 33(73)     |
| **Tier 3** | 5(178)   | 3(130)    | 5(94)     | 4(162)   | 14(334)  | 7(127)    | 4(99)   | 42(1,124)  |
| **Total**  | 12(207)  | 9(141)    | 12(107)   | 12(174)  | 33(365)  | 13(137)   | 9(105)  | 100(1,236) |

Note: Numbers in brackets represent the total number of Chinese universities in each category.

Due to limited data availability, we did not use random sampling for our data collection, which is a limitation for this study. When comparing the science and technology personnel (S&T personnel), number of international publications, research funding received, and the number of graduate students between the sample and the population (see Table 3), we found that the means of these indicators from the sample Tier 1 universities were very close to those means from all Tier 1 universities, while the means from the sample Tier 2 universities were only a little higher than the means from all Tier 2 universities. The Tier 3 sample seemed to include many top Tier 3 universities so that the sample means were much higher than the average of all Tier 3 universities. We also did the one-sample T-test (α=0.05) comparing the sample means with the population means to test whether the samples are representative. As Table 3 shows, we did not find any significant difference between sample and population in all four indicators in Tier 1 and Tier 2 and one indicator (S&T personnel) in Tier 3; significant difference was found between the Tier 3 sample and population in terms of the number of international publications, the research funding received, and the number of graduate. The T-test indicated that the Tier 1 and Tier 2 samples represented the population well while the Tier 3 sample was a little weak in this study.

---

[3] Some internal documents or information regarding the cash reward policies were provided privately by university staff.

**Table 3 Comparison of Stats between the Sample Universities and All Universities in Average (2014)**

|  | S&T personnel | International Publications | Research funding (USD: in millions) | Number of graduate students* |
|---|---|---|---|---|
| **Tier 1 Sample** | 5,182 | 3,071 | 205.97 | 16,176 |
| **All Tier 1** | 4,830 | 2,896 | 210.68 | 15,700 |
| **Diff (Tier 1)** | 0% | 0% | 0% | 0% |
| **Tier 2 Sample** | 2,228 | 807 | 62.16 | 7,937 |
| **All Tier 2** | 1,822 | 684 | 56.43 | 7,071 |
| **Diff (Tier 2)** | 0% | 0% | 0% | 0% |
| **Tier 3 Sample** | 1,045 | 290 | 19.05 | 3,348 |
| **All Tier 3** | 831 | 136 | 9.16 | 1,209 |
| **Diff (Tier 3)** | 0% | 30.9% | 32.8% | 105.1% |

Source: Ministry of Education of China (2014)
* Data regarding the numbers of graduate students were provided by Research centre for China Science Evaluation (RCCSE) at Wuhan University
Note: Diff in this table refers to the relative measure of hypothesized mean difference in the one-sample T-test (α=0.05). For example, 0% means that no significant difference between the sample mean and population mean while 30.9% means that the mean difference is equal to 30.9% of the population mean.

*Data Analysis*

Each cash reward policy contains various specifications about its criteria for the eligibility, amount, formula for calculation, and method of payment. It was difficult to compare different cash reward policies with different specifications. In order to compare the cash reward policies issued by different universities in different years, we selected some journals as examples and calculated the amounts of cash reward for a single research paper published in these journals according to different cash reward policies. The selected journals represent journals with different Journal Impact Factors (JIFs) and in different Journal Citation Report (JCR) Quartiles. For a good understanding of the comparison, we selected a list of nine popular journals that could be recognized by our readers, including four multidisciplinary science journals (the first 4) and five library and information science journals (the last 5) as shown in Table 4.

**Table 4 List of Journals Selected for the Comparison**

|  | Journal Impact Factor (5-year) | JCR Quartile (modified) |
|---|---|---|
| *Nature* | 41.458 | Q1 |
| *Science* | 34.921 | Q1 |
| *Proceedings of National Academy of Sciences (PNAS)* | 10.285 | Q1 |
| PLOS One | 3.535 | Q3 |
| *MIS Quarterly* | 9.510 | Q1 |
| *Journal of the Association for Information Science and Technology (JASIST)* | 2.762 | Q1 |
| Journal of Documentation | 1.480 | Q2 |
| *Library Hi Tech* | 0.741 | Q3 |
| *International Journal of Library and Information Science (LIBRI)* | 0.469 | Q4 |

Source: Journal Citation Report 2015

Both the Journal Impact Factor (JIF) of selected journals and the Journal Citation Report (JCR) Quartile in which these journals are located were used to calculate the amount of cash reward in most cash reward policies. Please note that we chose 5-year JIF instead of 2-year JIF because the former was used frequently by Chinese universities. Also, the JCR Quartile applied to the cash reward policies is not the original one with four equal quarters, but a modified one made by the Chinese Academy of Sciences. Compared to the original JCR Quartile grouping journals in each discipline into four equal quarters, the modified JCR Quartile use a pyramid hierarchy instead: only the top 5% of journals in each discipline are grouped into the Q1 while journals ranked in 5-20%, 20-50% and the bottom 50% are grouped into Q2, Q3 and Q4 in the modified JCR Quartile, respectively.

Results

A landscape of the cash-per-publication reward policy in China emerged as all 168 cash reward policies were analyzed. Chinese universities offer cash rewards that range from 30 to 165,000 USD for a single paper published in journals indexed by WoS, and the average reward amount has been increasing for the past 10 years. The results show us the overview of the cash-per-publication reward policies in terms of eligibility, amount, and their diversity and trends.

*Eligibility and Method*

WoS, which includes the Science Citation Index Expanded (SCIE), the Social Science Citation Index (SSCI), the Arts and Humanities Citation Index (AHCI) and the Conference Proceedings Citation Index (CPCI), plays a crucial role in China's cash reward policies. WoS data and the Journal Citation Report (JCR) are used as the eligibility criteria and the grade of the cash reward. Among all cash reward policies, only WoS papers are eligible for the cash reward, except that some universities offer small cash awards to papers indexed by *Engineering Index* (EI)[4]; WoS papers published in different journals may be awarded different amounts according to the journal's JIR and JCR quartile. Based on the analysis of these 168 cash reward policies, we grouped them into the following four major categories:

1. One-price reward (31): Universities pay the same amount to all WoS papers regardless of where these papers are published.
2. Original JIF-based reward (49): Universities award eligible papers different amounts based on the JIF of the journals in which these papers are published. Some universities assigned different grades to eligible journals on the basis of their JIF and pay more for papers published in journals with a high grade; some universities use the JIF as the multiplier to differentiate the cash reward (e.g. the amount of cash reward is equal to a basic amount times the JIF).
3. JCR Quartiles-based reward (99)[5]: Universities award eligible papers different amounts based on the modified JCR Quartile of the journals provided by Chinese Academy of Science that these papers published.
4. Citation-based reward (15): Universities award papers on the basis of the number of citations they received in a given citation window[6]. Some universities set up a threshold of the number of citations and award papers over the threshold; some universities use the Essential Science Indicators (ESI) (e.g., hot paper and highly cited paper) as the threshold and award these papers.

---
[4] Engineering Index is an engineering bibliographic database published by Elsevier. It indexes scientific literature pertaining to engineering materials.

Among these 168 cash reward policies, we found 31, 49, 99, and 15 policies that fell into these four categories, respectively. These numbers sum to over 168 because some cash reward policies were grouped into more than one category when universities apply multiple methods to awarding their international publications. We also found trends in the cash reward policies in effect in the following three stages from the late 1990s onwards, as shown in Figure 4. Both the one-price reward policies and the Original JIF-based reward policies were popular in the late 1990s and early 2000s, but their shares decreased when the JCR Quartile based policy was introduced. Since 2005, more and more Chinese universities have adopted the JCR Quartile based policy, which became the dominant policy from 2013 onwards.

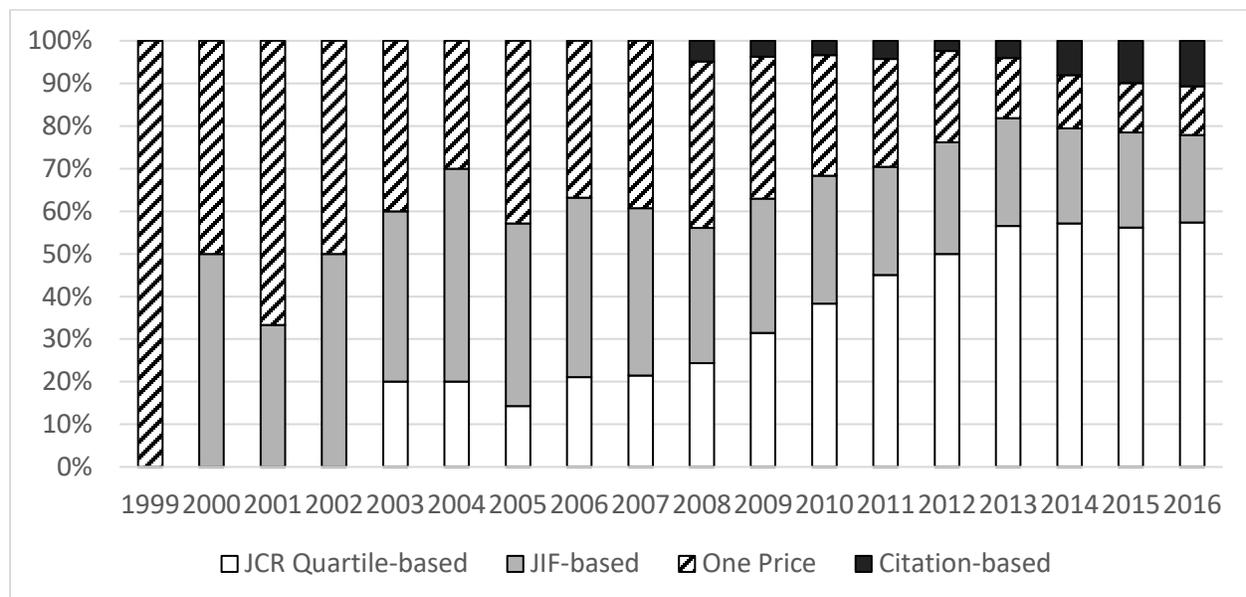

**Figure 4 Share of the Cash Reward Policies in Effect by Category (1999-2016)**

---

[5] Although the JCR Quartile is also based on the JIF, JCR Quartile provides a structural hierarchy that groups all journals into 4 categories, which the original JIF-based method does not.
[6] The citation window varies in different universities; some universities use a 5-year citation window while other universities use a 3-year citation window; some universities even count the citations at any time.

*Authorship*

The amount of individual cash rewards per WoS paper varies from 30 USD to 165,000 USD. Not all authors of a paper can claim cash rewards. In 118 out of 168 cash reward policies, universities only award cash to the first author; some universities even require that the awarded author must be both the first author and the corresponding author in 22 out of these 118 policies. Among 25 exceptional policies, universities award cash to non-first authors whose papers were published in particular prestigious journals (e.g., *Nature*, *Science*). Only 13 policies indicate that non-first authors may be awarded for all eligible publications, as they could get a discounted amount (e.g., half for the second author, a quarter for the third, etc.). In addition, there is no specific requirement for authorship in 12 out of 168 policies.

*Amount of the Cash Reward*

After analyzing 75 policies from 40 Chinese universities[7] that had the cash reward policy in effect between 2008 and 2016, we inferred a cash reward policy by calculating the average cash award for papers published in nine selected journals as described above. We also found that Chinese universities increased the amount of cash reward on average between 2008 and 2016 as shown in Table 5.

1. *Nature*, *Science*: Among most cash reward policies, publishing a paper in these two prestigious journals[8] would receive special treatment. Chinese universities offer the highest cash reward to *Nature* or *Science* papers. The author(s) may receive a prize up to 165,000 USD; some universities even announced that the amount of cash rewarded for a *Nature* or *Science* paper was *negotiable*. Indeed, the average amount of cash award for a *Nature* or *Science* paper increased 67% from 26,212 USD in 2008 to 43,783 USD in 2016.
2. *PNAS:* Although *Proceedings of National Academy of Sciences* is also a prestigious journal, it is not recognized by Chinese universities for special treatment. However, based on its JIF and JCR Quartile ranking, the average cash award for a *PNAS* paper is more than 3,000 USD, increasing slightly from 3,156 USD in 2008 to 3,513 USD in 2016.
3. *PLOS ONE:* Although *PLOS ONE* is ranked as a Q1 journal in the original JCR Quartile, it is categorized as a Q3 journal by the modified version of JCR provided by the Chinese Academy of Science. As a result, the amount awarded to a *PLOS ONE* paper is only around 1,000 USD, and this even declined from 1,096 USD in 2008 to 984 USD in 2016.
4. *MIS Quarterly, JASIST*: Both journals are ranked as Q1 journals in their category (Library and Information Science) by JCR. *MIS Quarterly*'s JIF is higher than *JASIST*'s, although the latter is recognized as the top journal in Library and Information Science, and so the average amount of cash awarded for a *MIS Quarterly* article is higher than that for a *JASIST* paper. From 2008 to 2016, the average amount of cash awarded for a *MIS Quarterly* paper slightly increased from 2,613 USD (2008) to 2,938 USD (2016), while the average amount of cash award to a *JASIST* paper increased 43% from 1,737 USD in 2008 to 2,488 USD in 2016.

---

[7] In order to keep the analysis consistent, 60 universities were excluded because their first cash reward policies were issued in 2009 and after.

[8] Some Chinese universities add *Cell* to this list of prestige journals; but most universities only recognize *Nature* and *Science*.

5. *Journal of Documentation*: As a journal ranked Q2 by the JCR, the average amount cash award to a paper published in *Journal of Documentation* is over 1,000 USD. The average amount increased from 1,082 USD in 2008 to 1,482 in 2016.
6. *Library Hi Tech, LIBRI*: Although both journals are indexed by WoS, they are respectively ranked as Q3 and Q4 journals in JCR because of their low JIF. These rankings are reflected in the cash awards: the average amount of cash awards to papers published in these two journals is below 800 USD. The average amount to LIBRI paper even decreased from 650 USD in 2008 to 484 USD in 2016.

**Table 5 Comparison of Average Amount of Cash Awards\* for a Paper Published in Selected Journals (2008-2016)**

|  | 2008 | 2009 | 2010 | 2011 | 2012 | 2013 | 2014 | 2015 | 2016 |
|---|---|---|---|---|---|---|---|---|---|
| *Nature, Science* | $26,212 | $26,006 | $25,781 | $25,365 | $33,990 | $36,658 | $38,908 | $43,783 | $43,783 |
| *PNAS* | $3,156 | $3,025 | $3,353 | $3,443 | $3,664 | $3,619 | $3,751 | $3,513 | $3,513 |
| *PLOS One* | $1,096 | $1,086 | $1,035 | $994 | $991 | $915 | $941 | $984 | $984 |
| *MIS Quarterly* | $2,613 | $2,570 | $2,553 | $2,654 | $2,876 | $2,861 | $2,992 | $2,938 | $2,938 |
| *JASIST* | $1,737 | $1,758 | $1,741 | $1,887 | $2,066 | $2,303 | $2,435 | $2,488 | $2,488 |
| *Journal of Documentation* | $1,082 | $1,087 | $1,042 | $1,111 | $1,167 | $1,265 | $1,329 | $1,408 | $1,408 |
| *Library Hi Tech* | $781 | $775 | $726 | $741 | $740 | $768 | $795 | $783 | $783 |
| *LIBRI* | $650 | $644 | $577 | $560 | $538 | $509 | $517 | $484 | $484 |

\* All the amounts are full amount (in USD) awarded to the first author

In summary, Chinese universities differentiate the amount of cash reward based on the JIF and JCR Quartile of journals in which the awarded papers are published. The average amount of cash award has increased over the past 10 years, except that the amount awarded to papers published in journals with low JIF has decreased. Publications in *Nature* and *Science* are awarded the largest amount of cash reward. This trend is also reflected by the change of 5 cash reward policies from Zhejiang University as shown in Table 6. The amount of cash awarded for publications in prestigious journals (i.e. *Nature, Science, PNAS*) increased while the amount for publications in other journals declined. Only the first author could receive cash rewards, except that non-first authors could get a discounted amount when publishing in *Nature* or *Science.*

**Table 6 Comparison of Cash Awards\* in 6 Cash Reward Policies from Zhejiang University**

|  | **2002** | **2005** | **2008** | **2010** | **2015** |
| --- | --- | --- | --- | --- | --- |
| *Nature, Science* | $6,000 | $30,000 | $30,000 | $30,000 | $45,000 |
| *PNAS* | $900 | $2,100 | $2,250 | $1,500 | $1,500 |
| *PLOS One* | $900 | $600 | $600 | $600 | $0 |
| *MIS Quarterly* | $900 | $750 | $900 | $900 | $1,050 |
| *JASIST* | $525 | $450 | $450 | $600 | $0 |
| *Journal of Documentation* | $525 | $450 | $450 | $0 | $0 |
| *Library Hi Tech* | $525 | $300 | $225 | $0 | $0 |
| *LIBRI* | $525 | $300 | $225 | $0 | $0 |
| *Eligible authorship* | 1$^{st}$ only | 1$^{st}$ only except Nature, Science | 1$^{st}$ only except Nature, Science | 1$^{st}$ only except Nature, Science | 1$^{st}$ only except Nature, Science |
| *Type of policies* | JIF | JIF | JIF | JIF | JCR Quartile |

\* All the amounts are full amount (in USD) awarded to the first author

*Difference by Tier and Region*

We also found that universities in different tiers have different preferences when choosing their cash reward policies. 14 out of 15 citation-based reward policies were issued by Tier 1 and Tier 2 universities, while 60% of the one-price reward policies were issued by Tier 3 universities. Such preferences also differ in universities from different regions. About 90% of universities in developed regions preferred either the original JIF-based reward policies or the JCR Quartile-based reward policies, while 60% of the one-price reward policies were favoured by universities in developing regions.

It was unexpected that Tier 3 universities would like to pay more for publications than Tier 1 and Tier 2 universities, despite having smaller budgets. As Table 7 shows, in 2016, Tier 3 universities paid double or even triple what Tier 1 and 2 universities did for a paper published in some journals. The average amounts of cash reward at Tier 2 universities for papers published in these journals are *between* the amounts paid by Tier 3 and Tier 1 universities, respectively. However, we did not find any significant difference in the average amount of cash reward among universities from different regions.

**Table 7 Average Amount of Cash Awards\* for a Paper Published in Selected Journals by Tier (2016)**

|  | *Nature, Science* | *PNAS* | *PLOS One* | *MIS Quarterly* | *JASIST* | *Journal of Documentation* | *Library Hi Tech* | *LIBRI* |
| --- | --- | --- | --- | --- | --- | --- | --- | --- |
| *Tier 1* | $38,846 | $2,704 | $401 | $1,924 | $1,465 | $817 | $283 | $216 |
| *Tier 2* | $53,823 | $4,113 | $783 | $3,251 | $2,695 | $1,377 | $679 | $434 |
| *Tier 3* | $63,187 | $5,488 | $1,661 | $5,150 | $3,902 | $2,102 | $1,172 | $642 |

\* All the amounts are full amount (in USD) awarded to the first author.

Discussion

Traditionally, the monetary reward incentive is used in business to reward employees with money for excellent job performance (Aguinis, Joo, & Gottfredson, 2013). Chinese universities apply this to awarding their scholars for research performance, thus promoting publication productivity. Considering the low annual salaries of Chinese scholars -- the average annual salary of university professors is around 8,600 USD, while the average basic salary of new hired professors is only about 3,100 USD (Altbach, 2012)[9] -- the amount of cash-per-publication reward is a huge incentive: the reward value for a JASIST paper is equal to a single year's salary for a newly hired professor while the cash award for a *Nature* or *Science* article is up to 20 times a university professors' average annual salary. The cash reward policy has been successful as China's international scientific publication has experienced a period of exponential increase in the past 20 years.

On the other hand, the monetary reward incentive also brings some negative effects. Chinese scholars may regard the monetary reward as an extrinsic rather than an intrinsic motivator (Aguinis et al., 2013; Kohn, 1993). In other words, they are driven to publish just for the monetary reward rather than disseminating knowledge and receiving the recognition defined by the reward system of science (Robert King Merton, 1973). For example, Professor Gao from Heilongjiang University published 279 papers in a single journal, *Acta Crystallographica Section E,* and received more than half of the total cash rewards given by Heilongjiang University between 2004 and 2009 (Lei & Lai, 2010). In this case, the monetary reward incentive creates a negative goal displacement effect (Frey et al., 2013; Osterloh & Frey, 2014). Prof. Gao's only research focus in these five years was to find new crystal structures in his lab and always report the results of this to the same journal, because he could accomplish the goal of winning the cash bonus in a short term as contrasted with receiving fewer awards by conducting long-term research projects (Lei & Lai, 2010). In addition, academic fraud in China, such as plagiarism, academic dishonesty, ghostwritten papers, fake peer review scandal, and so on also appeared in a growing number of publications (Hvistendahl, 2013). After searching the WoS, we found that the number of paper corrections authored by Chinese scholars increased from 2 in 1996 to 1,234 in 2016, a historic high.

This study also indicates the abusive use of bibliometric indicators in these cash reward policies. Although the Journal Impact Factor (JIF) is widely recognized to be a poor metric for evaluating the quality of individual papers (Archambault & Larivière, 2009; Lozano, Larivière, & Gingras, 2012; Seglen, 1997) , it is used in almost all cash reward policies as the golden rule[10] to assess the value of individual research. In addition, this study reveals that the WoS is the only data source accepted in these cash reward policies except that some universities offer small cash awards to papers indexed by *Engineering Index* (EI). Although Scopus could be an alternative to WoS in bibliometric studies (Norris & Oppenheim, 2007; Torres-Salinas, Lopez-Cózar, & Jiménez-Contreras, 2009) and indexes more Chinese journals (Mongeon & Paul-Hus, 2016), it is not recognized by Chinese universities. It also means that publications not indexed by WoS, including millions of papers published in Chinese journals, are almost ignored and excluded from the cash reward.

---

[9] In his book, Altbach used the purchasing power parity index instead of the pure salary based on the exchange rate for the comparison of professors' salary among 28 countries. It means that the annual salaries of Chinese scholars here (8,600 USD and 3,100 USD) represent the amount having the same purchase power as 8,600 USD and 3,100 USD in US respectively.

[10] Although the JCR Quartile is frequently used in the cash reward policies, the ranking of JCR Quartile is also on the basis of JIF.

We also found a positive trend among these cash reward policies. The focus of the cash reward policy changed from the quantity to the quality of the international publications when Chinese universities (especially Tier 1, 2 universities) had published an adequate number of WoS papers. This was why many one-price reward policies have been replaced by JCR Quartile-based or citation-based policies since 2008. In order to promote impact instead of quantity, these universities increased the amount of cash reward for papers published in Q1 and Q2 journals and decreased or stopped payment for papers published in Q3 and Q4 journals. A few universities even abandoned using the JIF and instead used the citation counts as the criterion for evaluating the quality of individual papers in citation-based reward policies. The hierarchical difference among universities was indicated in this trend change. When Tier 1 and Tier 2 universities reduced or cancelled the cash reward for papers published in journals with low JIF (e.g. Q3 and Q4 journals), some Tier 3 universities increased the amount for the same category papers. Indeed, Tier 3 universities have higher demand than Tier 1 and Tier 2 universities for both the quantity and the quality of international publications, so Tier 3 universities pay more cash for each individual paper.

Conclusion

In this study, after investigating 168 cash-per-publication reward policies from 100 Chinese universities, we described the landscape of the cash-per-publication reward policy in China and revealed its trends since the late 1990s. Chinese universities apply the monetary reward incentive used in business to promote scientific publication productivity, which lead to a radical increase in China's international scholarly publication. The cash-per-publication reward policy also produces the "Matthew Effect" (R. K. Merton, 1968) among university professors, as the amount of cash reward for publications is much higher than professors' annual salaries. Publications bring scholars not only cash rewards but also the possibility of future funding and promotion, which reveals the golden rule of academia in China: Publish or Impoverish.

This study revealed that monetary reward policies had been widely used to promote research productivity; these monetary reward policies might bring some negative effects when improving the research productivity, which was not systematically investigated by previous studies. We still know little about the potential impact of these monetary reward policies on research activities, which should be explored in the future. The landscape presented in this study could form a foundation for future studies that investigates the consequence and determinants of monetary reward policies.

Some limitations exist in this study. Due to limited data availability, we used convenience sampling, which may influence the sample representation as compared with sampling randomly, and the sampling fraction in Tier 3 universities is much lower than that of Tier 1 and Tier 2 universities. Social science and humanities is also not included in this study. We hope that future research could collect more data and overcome such limitations. Although this study presents a landscape of the cash-per-publication reward policies in China, we did not investigate if a correlation between the cash reward policy and the number of publications exists, which could be explored by future research.